\documentclass[5p,twocolumn,10pt,times]{elsarticle}
\usepackage{amsmath}
\usepackage{color}
\usepackage{hyperref}
\biboptions{authoryear}
\begin{document}
\baselineskip11pt

\begin{frontmatter}

\title{Embedded Spectral Descriptors: Learning the point-wise correspondence metric via Siamese neural networks}

\author[1]{Zhiyu Sun}
\author[1]{Yusen He}
\author[2]{Andrey Gritsenko}
\author[3]{Amaury Lendasse}
\author[1]{Stephen Baek\thanks{Corresponding author (\tt{stephen-baek@uiowa.edu})}}
\address[1]{Department of Industrial and Systems Engineering, University of Iowa, Iowa City, IA, United States}
\address[2]{Department of Electrical and Computer Engineering, Northeastern University, Boston, MA, United States}
\address[3]{Department of Information and Logistics Technology, University of Houston, Houston, TX, United States}

\begin{abstract} 
A robust and informative local shape descriptor plays an important role in mesh registration. In this regard, spectral descriptors that are based on the spectrum of the Laplace-Beltrami operator have been a popular subject of research for the last decade due to their advantageous properties, such as isometry invariance. Despite such, however, spectral descriptors often fail to give a correct similarity measure for non-isometric cases where the metric distortion between the models is large. Hence, they are not reliable for correspondence matching problems when the models are not isometric. In this paper, it is proposed a method to improve the similarity metric of spectral descriptors for correspondence matching problems. We embed a spectral shape descriptor into a different metric space where the Euclidean distance between the elements directly indicates the geometric dissimilarity. We design and train a Siamese neural network to find such an embedding, where the embedded descriptors are promoted to rearrange based on the geometric similarity. We demonstrate our approach can significantly enhance the performance of the conventional spectral descriptors by the simple augmentation achieved via the Siamese neural network in comparison to other state-of-the-art methods.
\end{abstract}


\end{frontmatter}

%\linenumbers

\section{Introduction}
\label{sec:1}
Point-to-point correspondence matching between surfaces is a fundamental problem in computer-aided design and computer graphics. The problem is critical for a wide range of applications such as shape interpolation \citep{Baek2015,Kilian:2007:GMS:1275808.1276457}, statistical shape analysis and modeling \citep{allen2003space,Baek2012}, geometry transfer \citep{sumner2004deformation}, and such. To this end, a well defined local shape descriptor and an accurate similarity metric can be greatly useful, as they provide a tool for quantifying the similarity between two points numerically.

For the last decade, a family of spectral descriptors \citep{Reuter2006,rustamov2007laplace,sun2009concise,aubry2011wave,bronstein2010scale} that utilizes the eigendecomposition of the Laplace-Beltrami operator  \citep{rosenberg1997} has drawn lots of attention among the researchers in relevant fields. This is because the spectrum of the Laplace-Beltrami operator has many advantegeous properties, such as isometry invariance, multiscaleness, parametrization independence, resistance to geometric/topological noise, and so on \citep{reuter2005laplace,sun2009concise}. In fact, the spectral descriptors in general show superior performances compared to other types of descriptors \citep{lian2013comparison}, especially for the tasks such as shape retrieval \citep{li2014spatially,bronstein2010scale}, mesh segmentation \citep{aubry2011pose,fang2011heat} and isometric matching \citep{ovsjanikov2010one}. However, despite of all the desirable properties, they often struggle in many correspondence matching tasks, especially when the disparity between the surfaces is relatively large. This is because the spectral descriptors, by their definition, are highly sensitive to the metric distortion. Therefore, even though they work generally well in isometric cases where the metric distortion is small (e.g., surface scans of the same individual with different postures), but not quite well in non-isometric cases (e.g., matching between a skinny person and an obese person).

In order to mitigate such limitations of the spectral descriptors, we introduce a novel approach to embed the spectral descriptors to a new metric space using artificial neural networks, in such a way that the Euclidean distance between the descriptors in the new embedding directly provides a reliable similarity metric for correspondence matching tasks. We design a Siamese neural network architecture, in which two parallel sets of neural networks sharing the same coefficients are used to train the suitable embedding. The Siamese NN is designed to find the optimal embedding of spectral descriptors by rearranging them in accordance of their correspondence among each other. In our experiment of detecting correspondence between human bodies, we demonstrate that the newly embedded shape descriptors through our Siamese NN, which we call \textit{embedded descriptors}, shows an enhanced performance in terms of the absolute number of correct matches and the distribution of the matches over entire shape.

\section{Related works}
\subsection{Spectral Descriptors}

Spectral shape analysis is a branch of computational geometry that analyzes digital geometry using the spectrum of a linear operator defined on a surface. Among a variety of choices, the Laplace-Beltrami operator \citep{rosenberg1997} that generalizes the Laplacian on Riemannian manifolds has gained significant highlights because of several desirable properties. One such property is the isometry invariance of its eigenvalues \citep{levy2006}. Since a lot of deformations in real-world can be characterized as an isometry or a near-isometry, the isometry invariance property of the eigenvalues of the Laplace-Beltrami operator gives advantage for many applications, including shape retrieval \citep{reuter2005laplace,jain2007spectral,li2014spatially,bronstein2010scale}, correspondence matching \citep{dubrovina2010matching,ovsjanikov2010one}, segmentation \citep{reuter2009discrete,aubry2011pose}, and etc.

Based upon this idea, \cite{Reuter2006} defined an ordered set of numerical signatures of a shape that captures the unique fingerprints of a given geometry. The signatures, which they named ``Shape DNA'', essentially are an ascending sequence of eigenvalues of the Laplace-Beltrami operator. They proved such an encoding of a geometry data provides an effective analytic tool for quantifying the geometric dissimilarity between different shapes in such a way that is proportional to the metric distortion.

Similarly, \cite{rustamov2007laplace} defined the global point signature (GPS), also by utilizing the eigendecomposition of the Laplace-Beltrami operator. He defined the GPS at each point on the surface as a vector value whose elements are the eigenfunctions of different modes scaled by the corresponding eigenvalues.

More notably, \cite{sun2009concise} proposed the heat kernel signature (HKS) physically analogous to the different heat diffusion characteristics depending on the geometric shape of surfaces. They introduced the heat kernel equation that formulates the isotropic heat diffusion process on a manifold surface by using the Laplace-Beltrami spectrum, and defined a shape signature as a collection of function values of the heat kernel. For each point on the surface, the heat kernel function is sampled at $n$ different time scales, forming an $n$-dimensional feature vector, to define the HKS. There have been several variants of HKS improving some of its properties, including works such as scale-invariant HKS \citep{bronstein2010scale} and persistent heat signature \citep{dey2010persistent}. The performance of the family of heat kernel descriptors has also been demonstrated in many related literatures \citep{ovsjanikov2010one, yoshiyasu2014symmetry, brandao2014partial,harik2017}.

In the similar spirit of HKS, \cite{aubry2011wave} proposed the wave kernel signature (WKS) that is based on a quantum mechanical characterization of the wave propagation on manifolds. The WKS represents the average probability of measuring a quantum particle at a specific location. This is achieved by solving the Shr\"{o}dinger's equation, whose solution is represented also by the Laplace-Beltrami spectrum in a similar form to the HKS.

Many relevant literature (see e.g., \cite{li2014spatially,lian2013comparison}) reports that spectral descriptors outperform other types of shape representation methods for shape retrieval tasks in general, since they are invariant to the isometry and are proportional to a deformation, or a metric change. Especially, signatures such as HKS and WKS are also known to be multiscale in a sense that they inherently capture both the local and global shape characteristics through different time scales.

However, the spectral descriptors are not quite suitable for the dense correspondence problems or deformable registration problems that involve large, non-isometric deformations. This is because, for non-isometric registration cases, the spectral descriptors tend to fail in recognizing the corresponding points between two models due to a large difference in local metrics. 

To mitigate such limitations, \cite{litman2014learning} introduced a machine learning approach for constructing a shape descriptor for the correspondence problems with large deformations. They generalized the well-known HKS and WKS equations into a generic form, and parameterized the generic form with a number of coefficients. They then proved that the Euclidean distance between the generic form of spectral descriptors are essentially the weighted distance with a metric tensor comprised of those coefficients. Therefore, by learning the optimal set of coefficients that generates the optimal metric maximizing the gap between true matching pairs and non-matching pairs, they derive an optimized spectral descriptor (OSD). The OSD made a significant contribution to the field by overcoming the limitations of conventional spectral descriptors to some extent. Similar approaches in attempts to learn informative descriptors that are used directly in the context of shape matching were adopted in several works \citep{windheuser2014optimal,corman2014supervised,rodola2014dense}. 

\subsection{Artificial Neural Networks in Correspondence Matching}
Artificial neural networks can find and generalize underlying rules and features of data on its own, through the observations on a training dataset, instead of using hard-coded or hand-crafted rules that are in general vulnerable to exceptions. This in fact is particularly useful for the tasks involving geometry data, such as computer vision, digital signal processing, and digital geometry processing, where features hold a highly sophisticated nature, and hence, manually coding sets of rules is extremely difficult or not necessarily available.

Indeed, in recent several years, there has been a phenomenal progress in artificial neural networks, with the emergence of convolutional neural networks (CNN), as nicely summarized in \cite{lecun2015deep}. In particular to the correspondence matching problem, works such as \cite{simo2015discriminative,kumar2016learning,balntas2016pn} employ CNNs to define shape descriptors on two-dimensional (2D) images. Instead of being analytically defined, shape descriptors in these works are derived from the CNNs through a number of observations on training images. The descriptors developed in such a way dramatically outperform the conventional image descriptors. 

Under the umbrella term, \textit{geometric deep learning}, there are many recent approaches that attempt to extend the tremendous success of CNNs to 3D geometric data analysis. \cite{wei2016dense} proposed using a CNN to train a feature descriptor for finding correspondences between non-rigid shapes. \cite{yi2016syncspeccnn} developed SyncSpecCNN by adapting the idea of graph convolutions on the spectral domain in order to achieve semantic segmentations of 3D shapes. \cite{guo20153d} firstly extracted traditional shape features by converting 3D data into vector, and learned a compact representation for mesh labeling. \cite{Masci_2015_ICCV_Workshops} proposed a geodesic convolutional neural networks (GCNN) as a generalization of CNN to Riemannian manifolds based on geodesic local patches. Using the polar coordinate system defined on each geodesic patch, the local features are extracted and passed through a cascade of filters and operators. The matching error can be minimized by optimizing the coefficients of the filters and the combination weights of the operators respectively. Building upon the GCNN, \cite{boscaini2016learning} constructed an anisotropic convolutional neural networks (ACNN) to further improve the matching performance of GCNN through the development of a patch operator using anisotropic kernels. More recently, \cite{verma2018feastnet} proposed a more advanced trainable parametrization scheme of the local neighborhood structure to improve kernels for correspondence learning.

\subsection{Contribution}
In this work, we propose a novel approach to learn shape descriptors for correspondence matching tasks between largely deformable shapes (non-isometric). Our major contribution is in the use of a data-driven approach to augment model-driven spectral descriptors for an application-specific use. Compared with the other state-of-the-art methods, our method shows a comparable performance to more sophisticated, computationally heavier approaches in terms of the overall matching accuracy.

%---------------------------------------------------------------------------
\section{Method}
\label{sec:2}
\subsection{Overview}
The goal of our method is to develop new shape descriptors by embedding the spectral descriptors into a new metric space such that the Euclidean distance between them directly provides a desirable similarity measure for deformable registration tasks. We first compute the spectral descriptors on each point of a 3D model. These spectral descriptors are then fed into our Siamese neural network as an input, in which the features are embedded into a new metric space. The output of the Siamese network is the similarity measure as well as a mapping from the space of original shape descriptors to the new metric space we desire.

\subsection{Spectral Descriptors}
\label{sec:spectral}
In this section, we describe the spectral descriptors that are to be used as inputs for our neural network. We used our own implementation for the computation of each type of the spectral descriptors and followed the standard parameters suggested in the original papers.

\paragraph{Global Points Signature \citep{rustamov2007laplace}} Given a point on a surface, the GPS at the point is defined as:
\begin{equation}
    \text{GPS}(x) = \left[
    \frac{\phi_1(x)}{\sqrt{\lambda_1}},\frac{\phi_2(x)}{\sqrt{\lambda_2}}, \ldots, \frac{\phi_n(x)}{\sqrt{\lambda_n}} \right]^\top,
\end{equation}
where $\lambda_k$ and $\phi_k$ are the $k$-th eigenvalue and eigenfunction of the Laplace-Beltrami operator defined on the surface respectively. An adequate number of eigenvalues suggested in \cite{rustamov2007laplace} is $n=25$. 

\paragraph{Heat Kernel Signature \citep{sun2009concise}} Given a surface, the heat flow on the surface can be approximated by the heat kernel function:
\begin{equation}
    H_t(x,y)=\sum_{k=0}^{\infty}\ e^{-\lambda_k t}\phi_k(x)\phi_k(y).
    \label{eq:heatkernel}
\end{equation}
Physically, the function returns the amount of heat diffused from a point $x$ to a point $y$ on the surface during a certain time $t$. From this, the HKS is defined as a series of heat kernel values $H_t(x,x)$ measured at discrete samples of time $t_1, t_2, \ldots, t_n$:
\begin{equation}
    \text{HKS}(x) = \left[ H_{t_1}(x,x),H_{t_2}(x,x),...,H_{t_n}(x,x) \right]^\top.
\end{equation}

\cite{sun2009concise} suggest using the first 300 eigenvalues and eigenvectors for the approximation of Equation~\ref{eq:heatkernel}. They also suggest uniformly sampling $n=100$ time samples in logarithmic scale over the time interval from $4\ln 10/\lambda_{300}$ to $4\ln 10/\lambda_2$.

\paragraph{Wave Kernel Signature \cite{aubry2011wave}} Given a surface, the propagation of a quantum particle on the surface is governed by the Schr\"{o}dinger's wave function, whose solution is given as follows:
\begin{equation}
    \psi_E(x,t) =\sum_{k=0}^{\infty}\ e^{i \lambda_k t}\phi_i(x)f_E(\lambda_k),
\end{equation}
where $i$ is the imaginary number and $f_E^2$ is an energy probability distribution with expectation value $E$. In practice, the energy probability distribution is approximated by the log-normal distribution, $e^{\frac{-(\rho-\ln \lambda_k)^2}{2\sigma^2}}$, where $\rho$ is the energy scale.

Here, the $l_2$ norm of the $\psi_E(x,t)$ physically has a meaning that the probability of measuring the particle at a point $x$ on the manifold at time $t$. The average probability is then achieved by integrating the norm over time:
\begin{equation}
    P_\rho(x) =\lim_{T\to\infty}\frac{1}{T}\int_{0}^{T}\|\psi_E(x,t)\|^2 dt =\sum_{k=0}^{\infty}\phi_k^2(x)f_E^2(\lambda_k).
    \label{eq:wavekernel}
\end{equation}

From this, WKS is defined as a series of the probability values in different energy scales $\rho_1, \rho_2, \hdots, \rho_n$:
\begin{equation}
    \text{WKS}(x) = \left[ P_{\rho_1}(x),P_{\rho_2}(x),...,P_{\rho_n}(x) \right]^\top.
\end{equation}

Similar to HKS, first 300 eigenvalues is used to approximate Equation~\ref{eq:wavekernel}, and the $n=100$ energy scale values are uniformly sampled over an interval from $\ln(\lambda_1)$ to $\ln(\lambda_{300})$, in \cite{aubry2011wave}.

\subsection{Embedding Space}
\label{sec:embedding}

Originally, spectral descriptors are in a canonical Euclidean $n$-space equipped with the typical Euclidean metric (i.e., $l_2$ norm). That is, each of the descriptors corresponds to a point in an $n$-dimensional Euclidean space, and the Euclidean distance between two points indicates the difference between the spectral descriptors, and thus, the geometric dissimilarity between the corresponding surface points.

Although the original $n$-dimensional spectral descriptors are informative about the geometric characteristics, they often contain a great deal of redundancy in description. Thus, one can visualize that the spectral descriptors lie on a lower dimensional manifold embedded in the Euclidean space. Hence, our goal is to utilize neural networks to characterize such a latent manifold space. As a first step, it is critical to determine the dimensionality of the new embedding space in such a way that it preserves most of geometric information encoded in the original spectral descriptors while keeping the dimensionality as concise as possible. We prove this hypothesis by conducting an analysis on the intrinsic dimensionality of the spectral descriptors.

The simplest way to accomplish this is to perform principal component analysis (PCA) on a set of spectral descriptors \citep{kirby2000geometric,wold1987principal}. First, we collect 10,000 random samples from our database. Then, we randomly pick a point from the samples and find $k$-nearest neighbors to the selected point. Next, we conduct PCA on the set of $k$-nearest neighbors to estimate the local tangent space around the selected point. Finally, we analyze the residual variances with respect to the number of principal components, and estimate the intrinsic dimension of the local tangent space from it. Technically, we find $d$ number of principal components that covers larger than 99\% of the total variance, that is $\sum_{i}^{d} \lambda_i \ge 0.99\sum_{i}^{n} \lambda_i$ where $\lambda_i$ is the eigenvalue associated with the $i$-th principal component. We repeat this process multiple times to statistically determine the intrinsic dimension of the spectral descriptors. More sophisticated methods, such as \cite{martin1979multivariate,tenenbaum2000global,roweis2000nonlinear}, may give better estimation on the intrinsic dimensionality of the data, we found them not critical from our experiments.

\begin{figure}
	\begin{center}
			\includegraphics[width=\columnwidth]{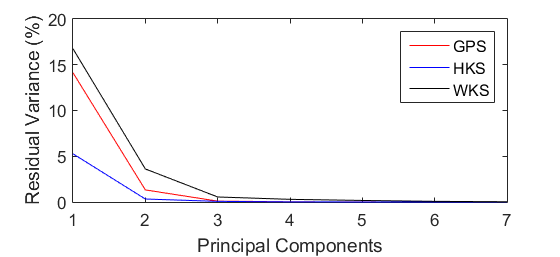}
	\end{center}
	\caption{Intrinsic dimensionality analysis. A significant drop of residual variance can be observed throughout the first five principal components. The trend is consistent when the number of nearest neighbors ($k$) is between 6 and 25. Hence, the intrinsic dimension of the spectral descriptors can be assumed to 5. We determine the dimension of the embedding space to be 15, which is three times bigger than the actual intrinsic dimension, in order to give some more flexibility of learning.}
	\label{fig:dimension}
\end{figure}

Figure~\ref{fig:dimension} shows graphs of the average residual variances for GPS, HKS, and WKS. As can be noticed from the figure, the residual variances drop significantly up to the first five or so components, and the later components are close to zero. Such a trend was consistent when we varied the number of nearest neighbors $k$ from 6 to 25. Interestingly, the same tendency was commonly observed across the different types of the spectral descriptors. We therefore conclude that the intrinsic dimension of the spectral descriptors is 5 and define the dimensionality of the embedding space to be 15, which is three times larger than the actual dimensionality to give enough degrees of freedom for distortion of the data manifold.

\subsection{Siamese Neural Network}
\label{network}
A ``Siamese'' architecture \citep{bromley1993signature} is an effective way of designing neural networks for comparative analysis. In Siamese network, two identical neural networks are placed in parallel, and the output layers of these networks are merged and are fed into another layers of neural network. The Siamese pairs share the same coefficients such that the weight values of the neurons and the biases are all identical between the pairs. The Siamese neural network is more advantageous than the other similar architectures. They can be trained with a fewer number of parameters since the weights are shared across the pair of neural networks. In addition, it is rational to use similar neural networks to process similar inputs (e.g., 3D models). Features with the same semantics extracted by deep learning techniques can be compared with each other easily \citep{koch2015siamese}.

\begin{figure*}
	\begin{center}
			\includegraphics[width=0.7\textwidth]{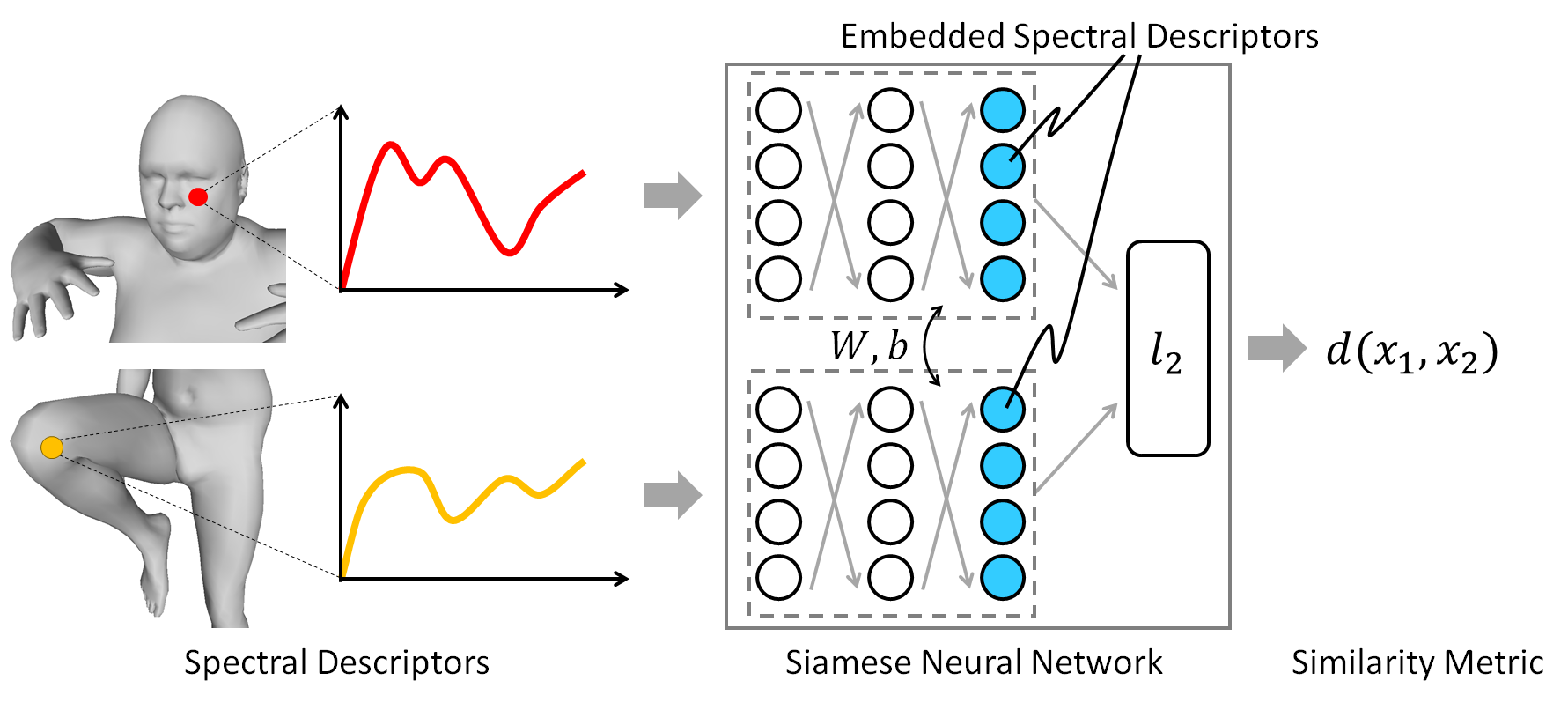}
	\end{center}
	\caption{A schematic diagram of the Siamese architecture used in this paper. Spectral descriptors computed at different points are fed into each branch of the Siamese network. The pair of the networks is identical, and shares the same coefficients (i.e., weights $W$, and bias $b$). The outputs of the Siamese pair, which mathematically are the spectral descriptors embedded into a different metric space, are then compared with Euclidean metric (or $l_2$ distance), which then gives a measure of similarity.}.
	\label{fig:siamese}
\end{figure*}

\begin{figure}
	\begin{center}
			\includegraphics[width=0.9\columnwidth]{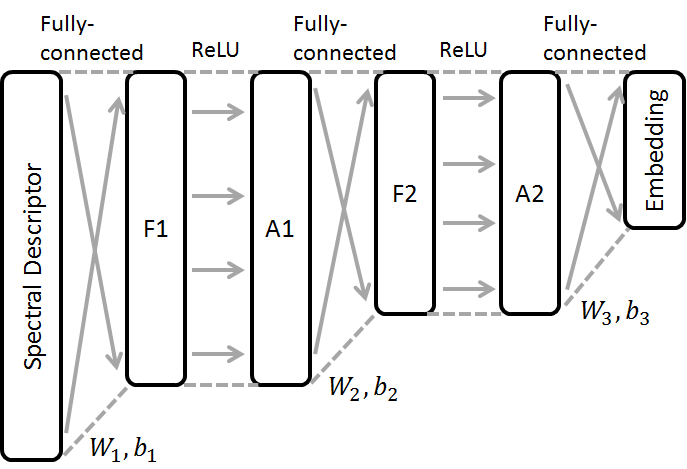}
	\end{center}
	\caption{The structure of the 5-layer network used in our method.}
	\label{fig:layers}
\end{figure}

In order to learn the optimal embedding of the spectral descriptors, we design a Siamese architecture as shown in Figure~\ref{fig:siamese}. First, each of the Siamese pairs takes a spectral descriptor defined at a given point as an input. Hence, the number of neurons on the input layer equals to the dimension of the spectral descriptor to be used. On the other hand, the output layer for each of the pairs is the final embedding of the spectral descriptors, and has a dimensionality of 15 as we discussed in Section~\ref{sec:embedding}. In-between the input and the output layers, there is a repetition of a fully-connected hidden layer and the rectified linear unit (ReLU) layer (Figure~\ref{fig:layers}). Finally, the two output layers from the both side of the Siamese pair are then merged and are fed into the main body, which is purposed to find the distance metric in the embedding space. It is a common practice to place another layers of neural network as a main body, however, in our case, we just set a simple unit for computing the Euclidean distance between the two embedded descriptors and transforming it to a similarity metric.

The most critical part of our design of the Siamese network is the hidden layers at each of the Siamese pairs. There are two main design parameters in concern for the design of the hidden layers: how many layers we require and; how many neurons we should have for each of the layers. Unfortunately, there are no known guideline that works well for most of the cases. In this reason, we conducted several sessions of experiments by varying the number of hidden layers and the number of neurons associated to each of it.

Empirically, we found that more than two fully-connected hidden layers tend to overfit the training data provided, and unnecessarily complicates the training process of the neural network. On the other hand, shallow networks with a single hidden layer showed reasonable performance in general, but not quite good as the two fully-connected hidden layer cases.

Given these, we also conducted a series of analysis to determine the adequate number of neurons for each of the fully-connected layers. Considering that HKS and WKS are 100-dimensional vectors and the output embedding is 15-dimension, we varied the number of neurons from 15 to 100 and trained the neural network for each of the cases. In the case of GPS, whose initial dimensionality is 25, we varied the values from 15 to 25. To evaluate the performance for different cases, we compared the loss (LSS), true-negative ratio (TNR), false-negative ratio (FNR), and the misclassification error rate (ERR), whose details will be discussed in Section~\ref{sec:evalcrit}.

\begin{figure*}
	\begin{center}
			\includegraphics[width=\textwidth]{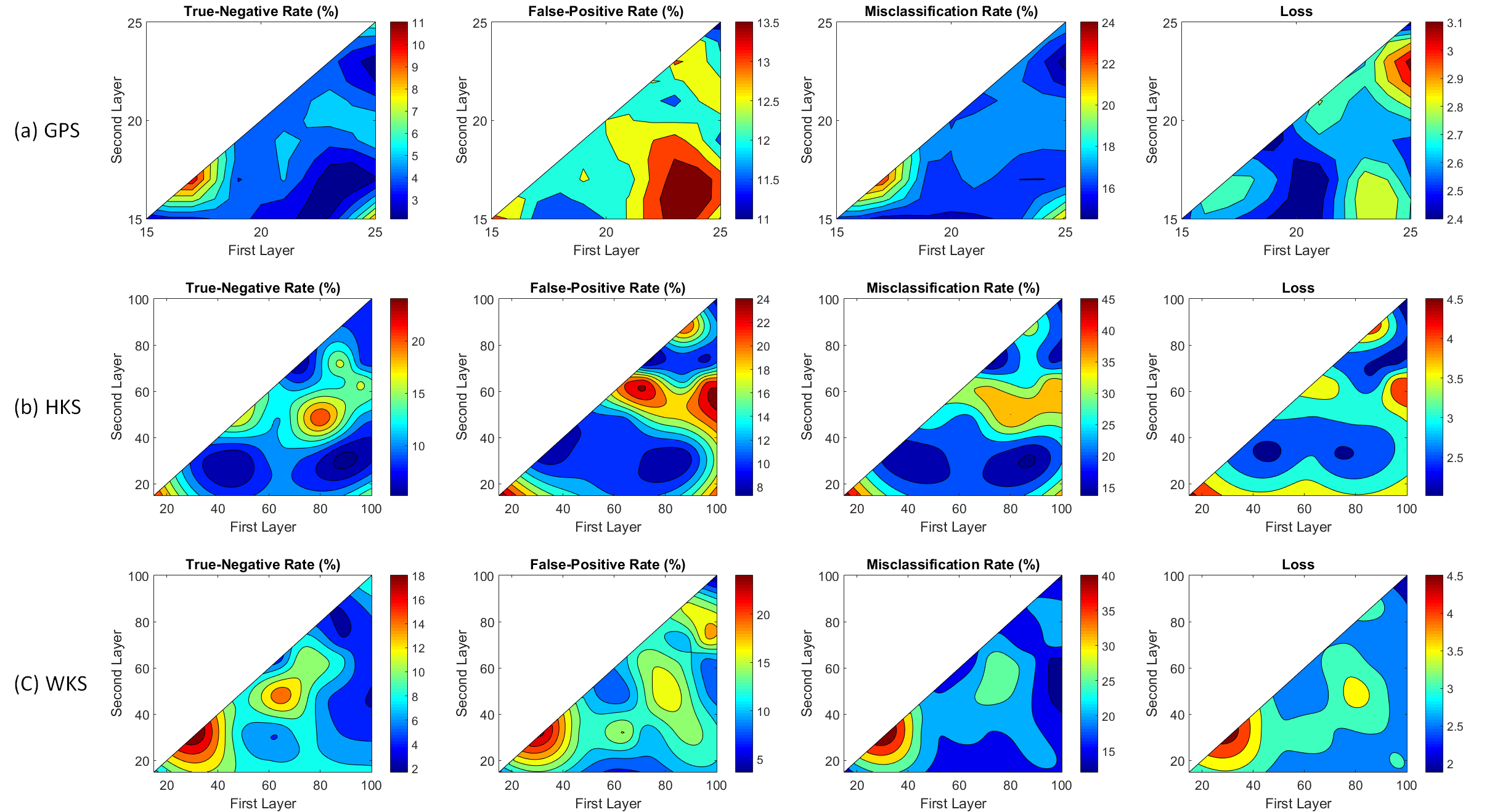}
	\end{center}
	\caption{Visualization of grid analysis for selecting an adequate number of neurons. Samples were collected every 2 units in each direction, and the values were interpolated for the visualization of the contour diagrams. From this, we choose 78 and 32 as the number of neurons for HKS and GPS, and 20 and 18 for GPS.}
	\label{fig:nnanal}
\end{figure*}

Figure~\ref{fig:nnanal} shows the result of our study. The horizontal axis represents the number of neurons in the first fully-connected hidden layer, and the vertical axis represents the second fully-connected hidden layer. Note that the upper triangle of the plot is empty because we excluded the cases where the next layers have the larger number of neurons than the previous layers. From such an analysis, we selected 78 and 32 as the adequate number of neurons for HKS and WKS, and 20 and 18 for GPS.

\subsection{Training}
\label{sec:training}

For the training of the neural network, we used the Dyna dataset \cite{Dyna} to generate our training samples. The Dyna dataset is composed of over 40,000 human body meshes of ten subjects spanning a range of body shapes under various body postures. Each mesh contains 6,980 vertices. The meshes are obtained by aligning the body scans of the 10 subjects. The body scans are collected by using a custom-built multi-camera active stereo system, which captures 14 assigned motions (e.g. running, jumping, shaking, etc.) at 60 frames per second. The system outputs 3D meshes of a size around 150,000 vertices in average. During the scanning, each subject wore tight pants and a bathing cap, and a sports bra for female subjects.

Such dense sets of vertices are then registered by conforming a template mesh to each of the scanned meshes. In this way, the topology of each mesh becomes compatible to each other, and the vertices with the same index get to correspond to each other geometrically. We exploit such by using them as the ground truth for the correspondence of a pair of vertices. On each of the meshes in Dyna database, we computed the spectral descriptors in a way that is described in Section~\ref{sec:spectral}, and stored their values as a data matrix.

For each batch at the training stage, we randomly select two models from the entire database, and pick 512 pairs of vertices at random. For each batch, the first half of the pairs are chosen from among the correct matching pairs, and the second half of the pairs are from the non-matching pairs.

The neural network then takes the pairs of spectral descriptors and updates its coefficients as a result of training. For the optimization, we found no significant difference between different types of optimization methods, but we found the Adam optimizer \cite{kingma2014adam} performs well in general. For the learning rate we used initially 0.015 and let it exponentially decay by a factor of 0.9999 for each of the training steps. The objective of the optimization is to maximize the margin between non-matching pairs. This can be achieved by minimizing the sum of the following error terms over all training samples $k$.

\begin{equation}
\begin{aligned}
    E(f_k, g_k)& =  y_k\| D(f_k)-D(g_k) \|^2 \\
    &+ (1-y_k)\max \left(0, C-\|D(f_k)-D(g_k)\|^2 \right),
\end{aligned}
\label{eq:error}
\end{equation}
where $(f_k, g_k)$ are the $k$-th pair, $y_k$ is a Boolean value indicating the correspondence, $C>0$ is the margin. $D(\cdot)$ is an embedding derived from the Siamese branch, and $\max (\cdot, \cdot)$ is a function comparing two values and returning the larger number. Theoretically, the objective function must try to separate non-matching pairs apart as far as possible in the feature space. Practically, the margin is set to $C=5$ instead of an unreasonably large number to achieve numerical stability.

Here, one practical consideration that needs to be done for the selection of the Boolean value $y_k$ is that even the non-matching pairs might have a highly similar geometry. For instance, a point on the left thumb would probably be highly similar in geometry with a point on the right thumb. However, they are technically non-matching pair. If such cases happen in the training dataset, it may confuse the neural network since it is a sort of conflicting example. In this reason, instead of setting $y_k$ strictly either 0 or 1, we set some random value between 0 to 0.2 for the non matching pairs. The correct matching pairs are still kept to $y_k=1$.

\subsection{Evaluation criteria}
\label{sec:evalcrit}

For the quantitative measurement of the trained performance of our method, we use following criteria:

\paragraph{Loss} Loss is the sum of error terms in Equation~\ref{eq:error} over test dataset. The loss is an indicator of how well the non-matching pairs are separated from each other in the embedding space.

\paragraph{True-Negative Rate (TNR)} TNR is the percentage of error that the correct matching pairs are classified as non-matching pairs. For the threshold of classification, we use the half of the margin $C$ in Equation~\ref{eq:error}.

\paragraph{False-Positive Rate (FPR)} FPR is the percentage of error that the non-matching pairs are classified as correct matching pairs. We use the same threshold with the above case. FPR illustrates the percentage of non-matching pairs are within the margin of $0.5C$.

\paragraph{Misclassification Rate (ERR)} ERR is the percentage of error that the pairs are classified to other than their ground truth. Conceptually, it is the union of TNR and FPR, and it is calculated as the sum of these two quantities.

%----------------------------------------------------------------------------------------------------
\section{Result}
\label{sec:result}

\subsection{Matching Performance}
\label{sec:3}
From the Dyna dataset, we randomly selected a pair of models from two different individuals that were not used for training of the Siamese DNN. We computed the spectral descriptors as well as the proposed descriptors on the models and matched their vertices one to the other through the nearest neighbor search using the simple Euclidean distance. We have rejected non-matching pairs, i.e., the pairs with the Euclidean distance in the embedded space over the threshold value $0.5C$. We also rejected the pairs with the geodesic distortion more than 5\% of the shape diameter in the spatial domain, similar to the testing criteria in \cite{litman2014learning}. We then counted the number of remaining matches, which could be considered as the correct matching pairs. We did not use any sophisticated algorithm for the calculation of the matching pairs, such as the ones presented in \cite{litman2014learning,leordeanu2005spectral}. Instead, we directly compared the Euclidean distances among the descriptors such that a more straight forward comparison could be done on the quality of the metric.

\begin{figure*}
	\begin{center}
			\includegraphics[width=\textwidth]{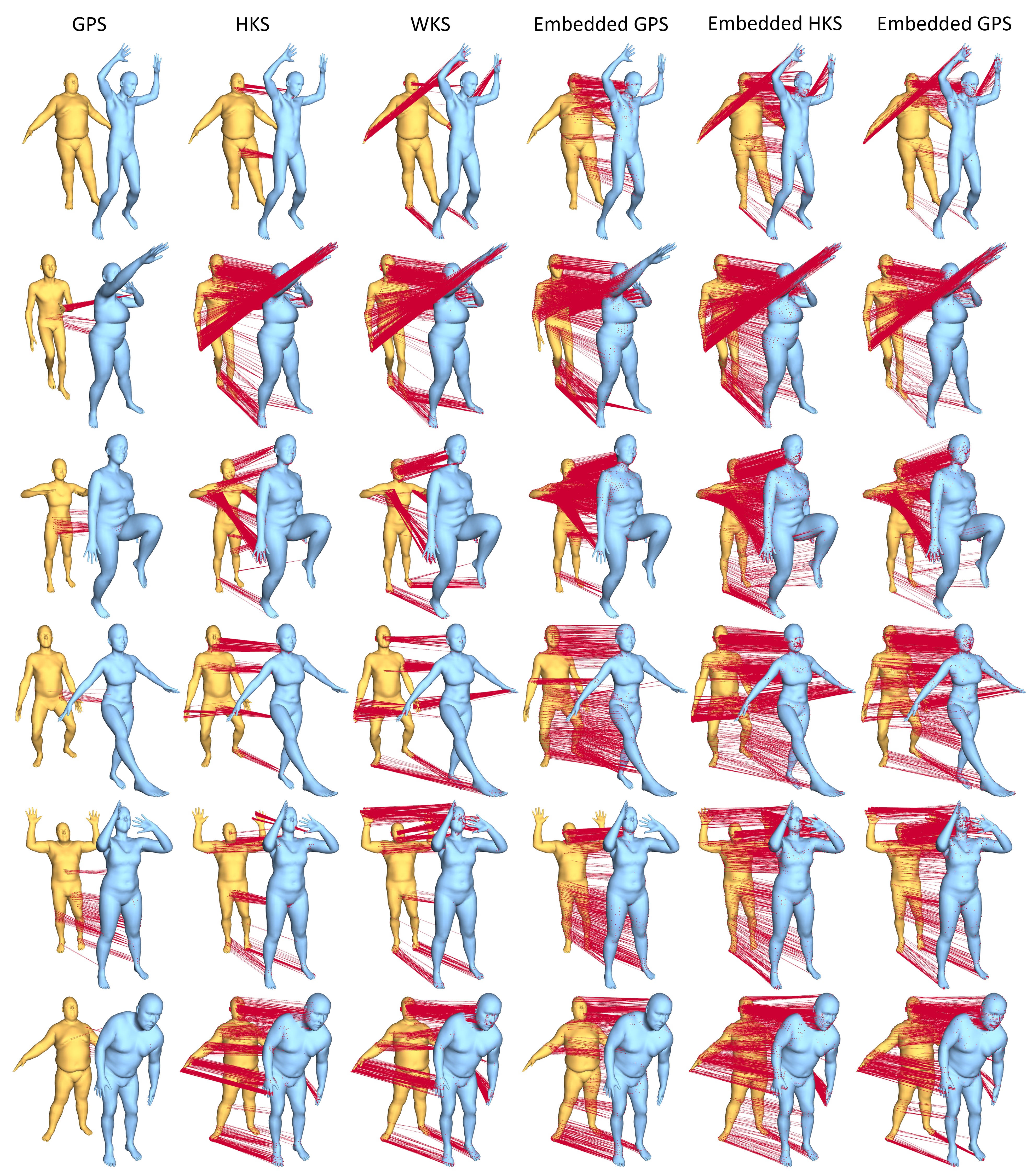}
	\end{center}
	\caption{Visualization of the matches within the geodesic distortion of 5\% of the shape diameter. Each row shows a comparison between the GPS, HKS, and WKS, and their embeddings. Different models in different rows are randomly picked from the test dataset for the visualization. For a fair comparison, no sophisticated algorithm such as \cite{leordeanu2005spectral} is used, but a simple nearest neighbor search based on the Euclidean distances between the shape descriptors.}
	\label{fig:bestmatch}
\end{figure*}

Figure~\ref{fig:bestmatch} and Table~\ref{tbl:DynaTest} show the result of such analysis. Overall, it is observed a significant improvement with the proposed method in terms of the absolute numbers of correct matching pair. Among them, GPS showed the largest improvement in average, followed by HKS and WKS. In overall matching performance, embedded HKS showed the most desirable result, followed by embedded WKS and embedded GPS.

Note, even for the cases where the improvement in absolute number was not so significant or even negative, the quality of matching still was improved in a sense that the pairs in deep signature results cover larger area than the original descriptors. For instance, the second row in Table~\ref{tbl:DynaTest} shows a relatively small improvement in HKS (2,209 $\rightarrow$ 2,300), and even a decrease of numbers in WKS (2,124 $\rightarrow$ 1,325) in terms of the absolute number of correct matching pairs. However, as can be observed from the corresponding images in Figure~\ref{fig:bestmatch}, the quality of matching is much better in embedded HKS and embedded WKS, such that the matching pairs are spread well over the larger area (see e.g., areas around breast and belly). This is, in fact, because some of the body parts such as hands, feet, and faces, are near-isometry across the models. In this reason, the matching results of the original spectral descriptors are mostly concentrated around those body parts, whereas the matching results of the deep spectral descriptors are distributed widely over the entire body area, even including highly non-isometric areas.

\begin{center}
\begin{table*}
\begin{tabular}{l|l|r|r|r|r|r|r}
    Source model & Target model & GPS & HKS & WKS & EGPS & EHKS & EWKS \\
    \hline
    50002\_one\_leg\_jump\textbackslash00455 & 50027\_jumping\_jacks\textbackslash00228 & 0 & 211 & 669 & 631 & 1,585 & 706 \\
    50009\_running\_on\_spot\textbackslash00165 & 50004\_punching\textbackslash00170 & 147 & 2,209 & 2,124 & 2,374 & 2,300 & 1,325 \\
    50020\_chicken\_wings\textbackslash00090 & 50022\_one\_leg\_jump\textbackslash00266 & 84 & 1,145 & 1,328 & 3,478 & 2,654 & 1,595 \\
    50026\_light\_hopping\_loose\textbackslash00168 & 50020\_knees\textbackslash00374 & 25 & 627 & 1,072 & 1,072 & 2,229 & 1,794 \\
    50007\_jiggle\_on\_toes\textbackslash00233 & 50022\_shake\_arms\textbackslash00302 & 206 & 455 & 1,162 & 998 & 1,637 & 1,684 \\
    50002\_hips\textbackslash00515 & 50007\_shake\_shoulders\textbackslash00214 & 17 & 1,474 & 1,943 & 921 & 1,975 & 1,565
\end{tabular}

\caption {Statistics of correct matching pairs for the results presented in Figure~\ref{fig:bestmatch}. Each of the rows in the table corresponds to each of the rows in Figure~\ref{fig:bestmatch} in the same order. The left two entries of each row show the id number of an individual and the name and the frame number of a motion, following the same naming convention in Dyna database. Different id numbers indicate different body shapes, and different motion names and frame numbers indicate different body postures. The rest of the entries show the number of correct matching pairs for each of the methods.}
\label{tbl:DynaTest}
\end{table*}
\end{center}

\subsection{Comparison with other methods}

\label{sec:4}
In addition to the aforementioned performance evaluation, we further verify the performance of our method by comparing with the OSD \citep{litman2014learning}, the GCNN \citep{Masci_2015_ICCV_Workshops} and the ACNN \citep{boscaini2016learning}. For such evaluation, we formulate a dataset containing $100$ different models by randomly picking $10$ meshes of each subject over all assigned motions across the total $10$ subjects in the Dyna dataset. From the formulated dataset, we used $60$ mesh models of 6 different subjects for training. The rest $40$ mesh models of the other $4$ subjects were kept for testing. In order to make the comparison fair, we used the $100$ dimensional HKS computed as described in Section~\ref{sec:spectral} as an identical network input for GCNN, ACNN and our Siamese neural network. Under such setting, we have a clear guide to compare the proposed methods by seeing how much extent they can improve some existing shape descriptor for a specific task (correspondence matching for deformable shapes).

To drill-down to details, we conducted four groups of analyses to evaluate the correspondence matching performance of each method. For all groups, the source models were selected from the mesh models belongs to the same subject ``50020'' which has the smallest body size among the four subjects in our test dataset. For the first group, the target models were selected from the same subject ``50002'', but with different postures corresponding to each source one. From the second to the fourth group, the target models are selected from the rest three subjects (``50021'', ``50025'', ``50002'') in an order of increasing body shape disparities compared with the subject ``50020'', respectively. Each of the group contains $10$ different pairs of models (source and target) that are randomly selected under the restriction of aforementioned manner.

\begin{figure}
	\begin{center}
			\includegraphics[width=\columnwidth]{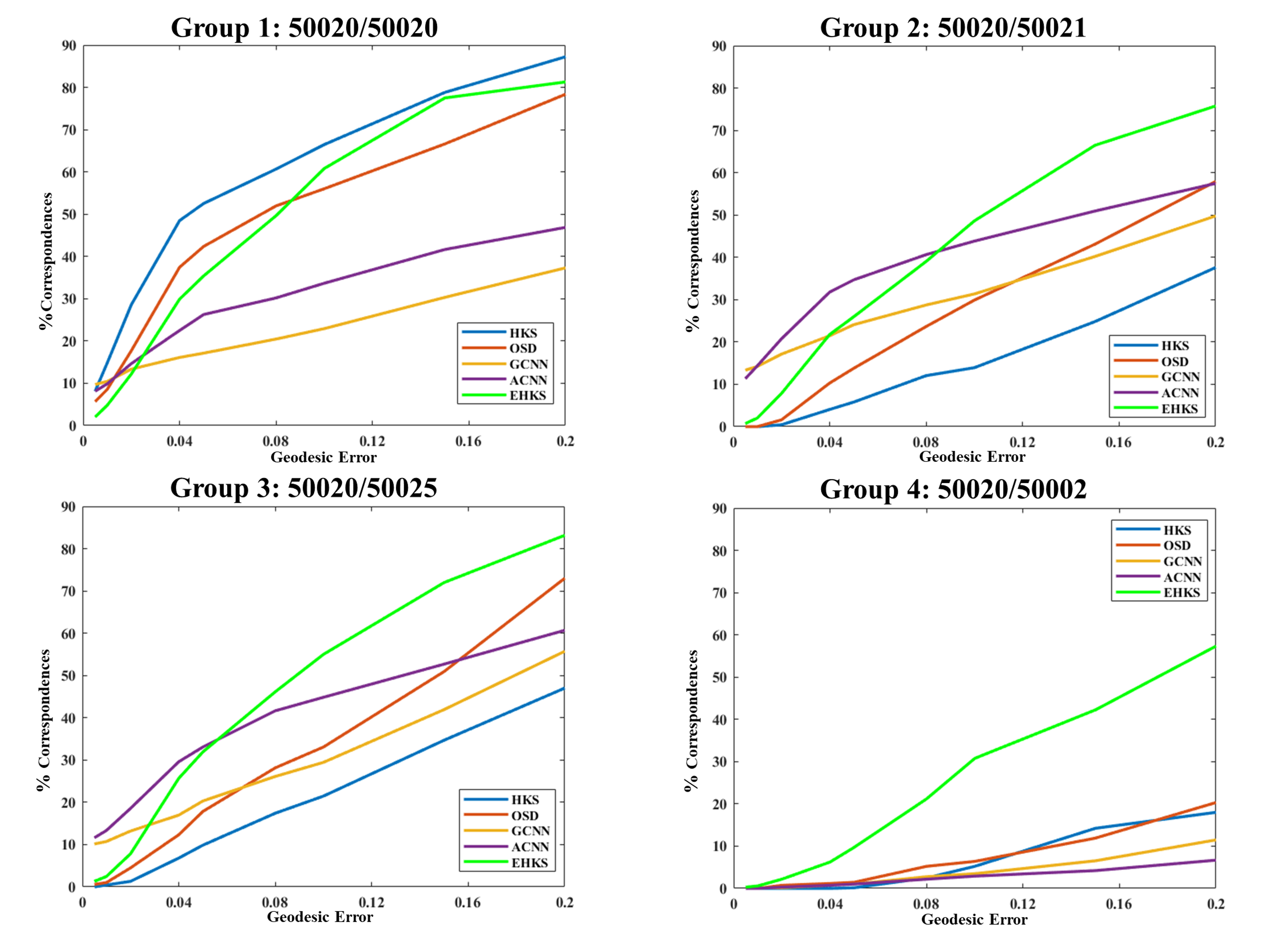}
	\end{center}
	\caption{Correspondence matching performance for four groups of analyses.}
	\label{fig:geodesic_error}
\end{figure}

\begin{figure*}
	\begin{center}
			\includegraphics[width=\textwidth]{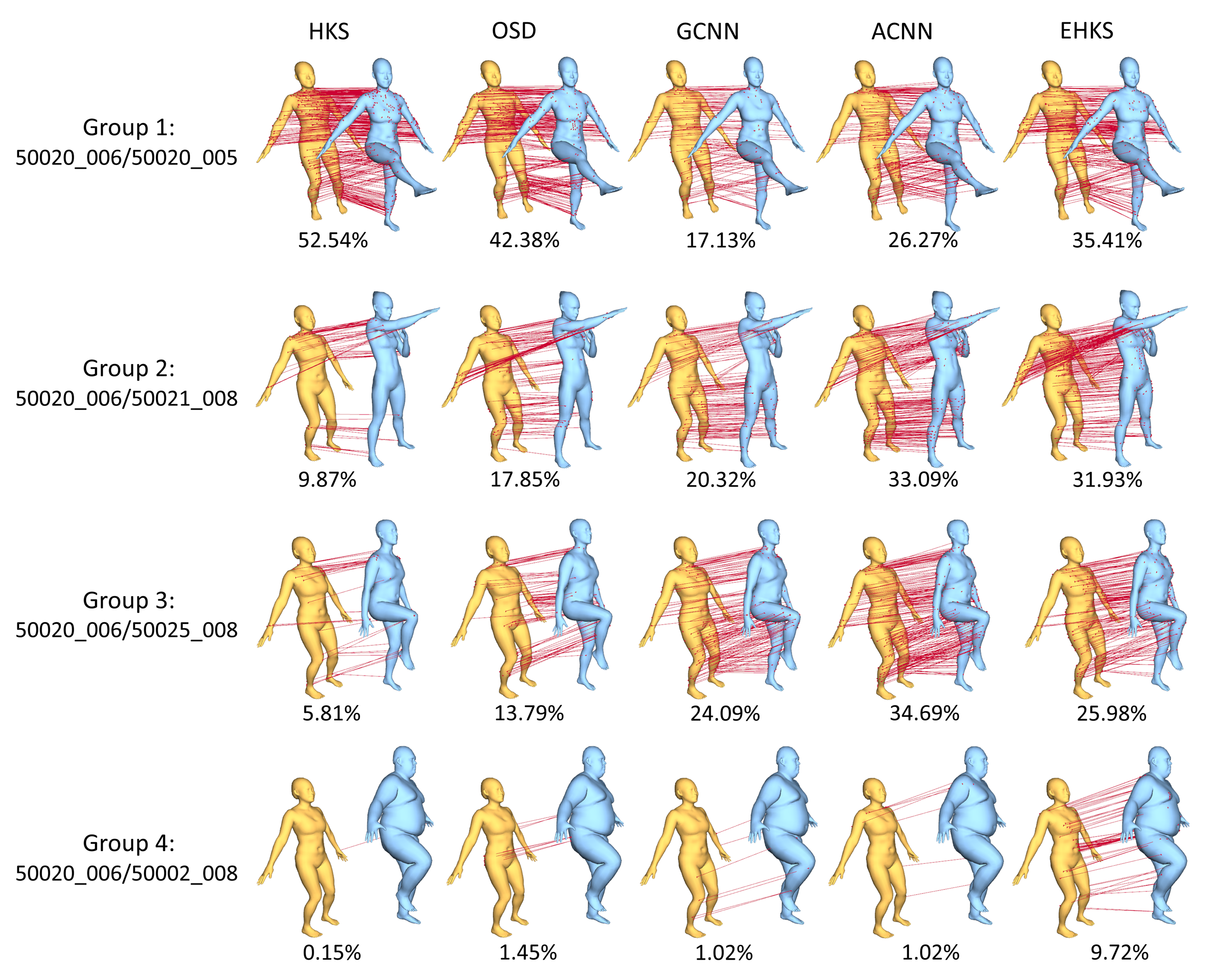}
	\end{center}
	\caption{Comparison among HKS, OSD, GCNN, ACNN and embedded HKS. Each row corresponds to a representative pair of models in each of the group presented in Table~\ref{tbl:DynaTest2}. For each of the pair, source model is rendered in yellow and target model in blue.}
	\label{fig:bestmatch3}
\end{figure*}

First we evaluated the matching performance for different methods over the four groups using the Princeton benchmark \cite{kim2011blended}. The matching performance is measured by the percentage of correctly predicted correspondence within a geodesic disk around the ground-truth point. The results were plotted in Figure \ref{fig:geodesic_error} with varying radii of the geodesic disk from 0\% to 20\% of shape diameter. For better visualization, in addition, we randomly sampled 10\% of all vertices uniformly distributed over the entire mesh. In the Dyna dataset, some of the body segments such as head, hands, and feet were rigid and thus remain unchanged across individuals due to the nature of how the dataset had been generated (see \cite{Dyna} for the detail). Thus, these areas were excluded from the evaluation for more precise comparison between different methods. For each pair of the source and target models, matching accuracy was calculated as the percentage value of counted correct matching pairs of vertices (within the geodesic distortion of 5\% of the shape diameter in the spatial domain.) over the total number of sampled vertices. Table~\ref{tbl:DynaTest2} summarizes the correspondences matching performance of each proposed method for the four groups of analysis as described. Figure~\ref{fig:bestmatch3} visualizes the correct matches of a representative pair for each of the group, with the corresponding matching accuracy shown in the figure. The visualization of additional pairs in each of the group is provided in the supplementary materials.

From Figure \ref{fig:bestmatch3}, it can be visually recognized that the body shape disparities between the paired subjects gradually increased from Group $1$ to Group $4$. For the Group $1$, while the deformation between source and target models is isometric (same subject with different posture), the HKS has the best performance in terms of the correspondence matching accuracy. As the amount of metric distortion increases (Groups 2, 3,\& 4), the matching accuracy of the HKS significantly drops, which is consistent with our prior knowledge according to the limitation of spectral descriptors. From here, the methods (OSD, GCNN, ACNN and embedded HKS) start to demonstrate their values in improving the HKS for large deformation (non-isometric) cases. For Group 2 and Group 3, the ACNN performed the best in terms of the overall matching accuracy, followed by our method (embedded HKS) that showed a comparable performance with ACNN. For Group 4, while the amount of metric distortion between models became tremendous (highly non-isometric), our method significantly outperformed all other methods in comparison. It is also noteworthy that our method showed a consistent level of the relative standard deviation (RSD) across all the groups and the smallest RSD in each of the group across all methods, which indicates a more reliable and robust performance of our method.

\begin{center}
\begin{table*}
\begin{tabular}{l|l|rr|rr|rr|rr|rr}
     & & \multicolumn{2}{c|}{HKS} & \multicolumn{2}{c|}{OSD} & \multicolumn{2}{c|}{GCNN} & \multicolumn{2}{c|}{ACNN} & \multicolumn{2}{c}{EHKS} \\
    Group ID & Subjects (source\textbackslash target) & Mean & RSD & Mean & RSD & Mean & RSD & Mean & RSD & Mean & RSD \\
    \hline
     1 & 50020\textbackslash 50020 & 54.18\% & 0.15 & 42.84\% & 0.22 & 15.69\% & 0.16 & 26.46\% & 0.10 & 34.73\% & 0.05 \\
     2 & 50020\textbackslash 50021 & 10.71\% & 0.20 & 20.13\% & 0.18 & 20.76\% & 0.21 & 33.03\% & 0.11 & 28.91\% & 0.08 \\
     3 & 50020\textbackslash 50025 & 5.88\% & 0.28 & 17.18\% & 0.14 & 22.25\% & 0.17 & 31.93\% & 0.09 & 26.15\% & 0.08 \\
     4 & 50020\textbackslash 50002 & 0.12\% & 0.53 & 1.80\% & 0.27 & 0.54\% & 0.52 & 1.05\% & 0.28 & 10.12\% & 0.10
\end{tabular}
\caption {Statistics for four groups of analyses. The mean matching accuracy and the relative standard deviation (RSD) over the $10$ pairs of selected models in each of the group are presented.}
\label{tbl:DynaTest2}
\end{table*}
\end{center}

\subsection{Geometric \& Topological Noise}
\begin{figure}
	\begin{center}
			\includegraphics[width=\columnwidth]{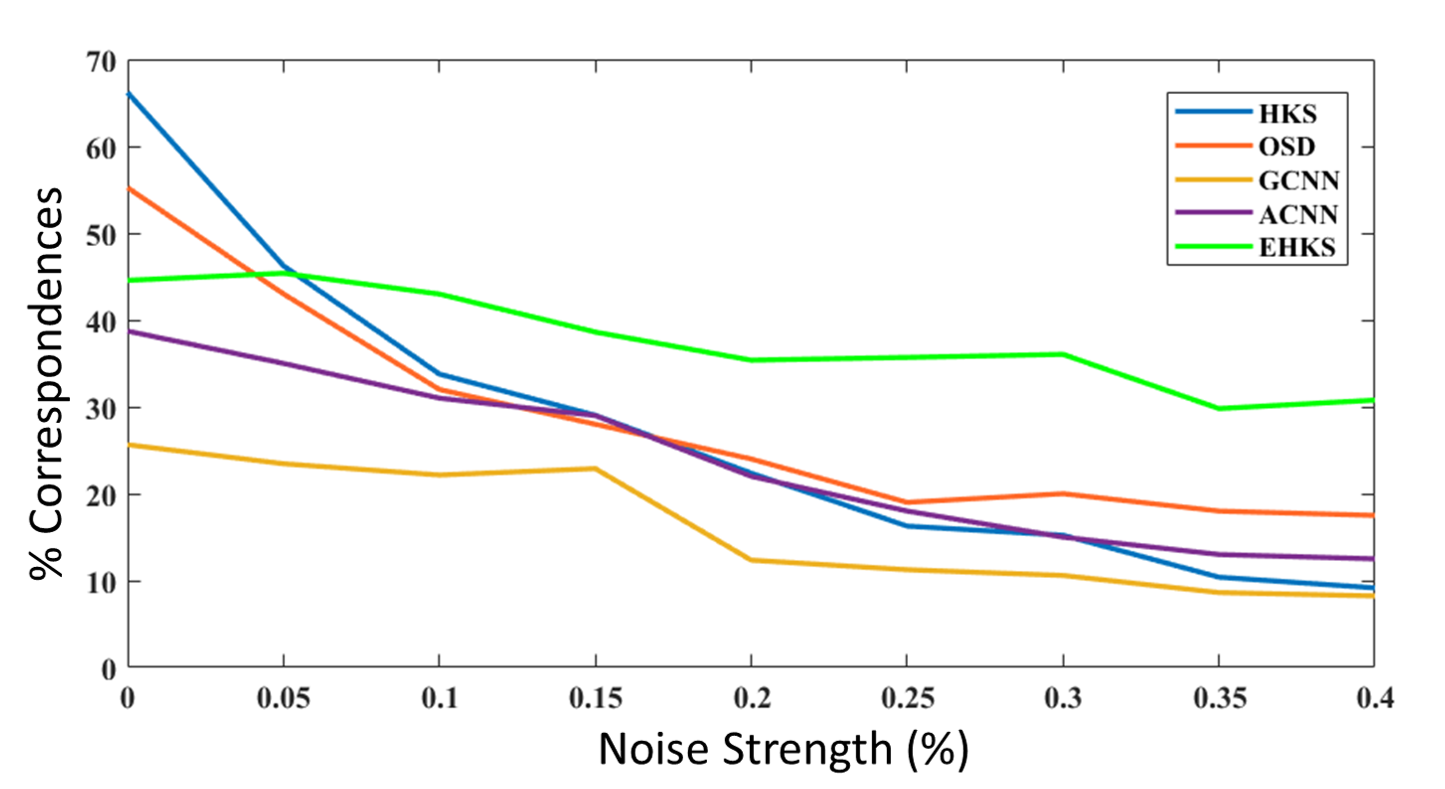}
	\end{center}
	\caption{Performance of correspondence matching in presence of noise.}
	\label{fig:partial matching error}
\end{figure}

\begin{figure*}
	\begin{center}
			\includegraphics[width=\textwidth]{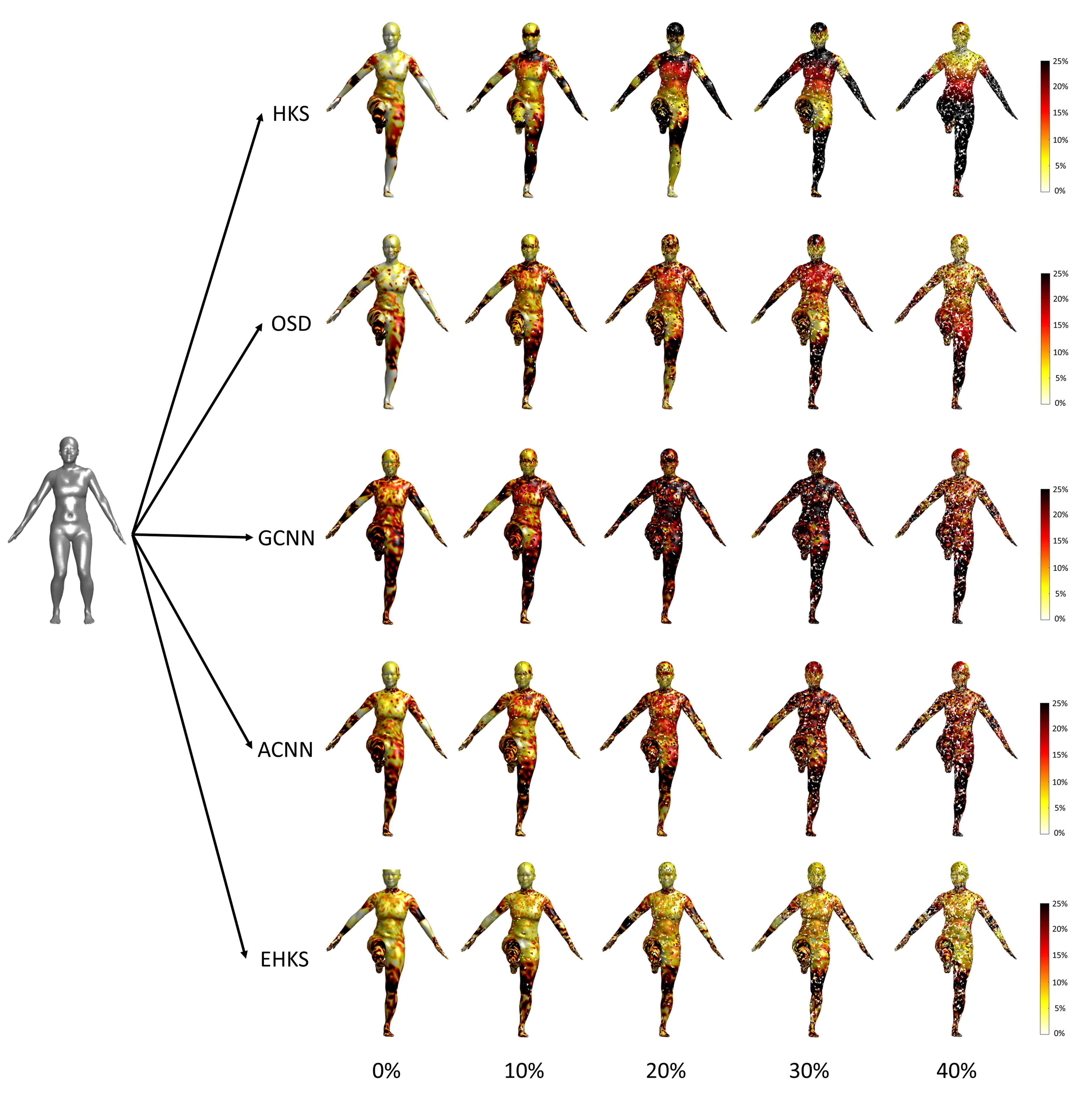}
	\end{center}
	\caption{Matching error distribution under noise. Each column represents different strength of noise. The color indicates scale of geodesic error from 0\% to 25\% of the shape diameter.}
	\label{fig:partial matching vis}
\end{figure*}

We further investigate the robustness of our method for correspondence matching on partial and noisy models. We erode a model randomly picked from the dataset by generating geometric and topological noises (i.e., random perturbation in mesh position and connectivity) at different levels. These noisy models at different levels are then matched to a reference model. Performance of HKS and embedded HKS was analyzed and displayed in Figure \ref{fig:partial matching error}. As the strength of error increases, the performance of HKS deteriorates significantly, while the embedded HKS remains quite robust. Figure \ref{fig:partial matching vis} shows a visual comparison of the point-wise geodesic correspondence error between these methods.

%---------------------------------------------------------------------------------------------------
\section{Conclusion}
\label{sec:5}
In this paper, we proposed an approach to learn the similarity metric between the spectral descriptors for correspondence matching tasks. We designed a Siamese neural network architecture, based on the analysis of its design hyperparameters. The neural network designed in this paper was able to find an optimal embedding of the spectral descriptors in a new metric space, in which a direct comparison based on the Euclidean distance already gives an good point-wise matching result between non-isometric models. The method was verified on a dataset containing a variety of body shapes and postures, where it showed significant improvement of the original spectral descriptors in non-isometric cases. In the task of improving a conventional spectral descriptor for detecting correspondence between non-isometric shapes, our method showed a comparable performance in terms of the overall matching accuracy with the best robustness compared with the other state-of-the-art methods. For highly non-isometric cases, our method significantly outperformed all the other methods.

It should be noted that more recent methods in geometric deep learning could certainly achieve higher performance. However, those methods are typically data-hungry, requiring a massive amount of data. Furthermore, implementation of geometric convolution is complicated and requires a great deal of work for preprocessing, to define, for example, geodesic polar coordinates around each of the vertices. On the contrary, the method proposed in this paper can be achieved with a simple implementation and a smaller amount of data, while demonstrating a comparable performance.

Meanwhile, one may wonder if concatenating different spectral descriptors all together and applying the same Siamese embedding would produce any better result. Although we have not reported in the results, we failed to observe any noticeable performance improvement led by the concatenation of GPS, HKS, and WKS. This could be due to the substantial overlap of information across different descriptors, as they all are originated from the eigenvalues and eigenfunctions of the Laplace-Beltrami operator.

As a future work, it would be of interest to explore possible ways of improving our method to achieve scale-invariance. Current formulation is dependent upon scaling of the model. Furthermore, transferring the learned embedding from one training set to another would also be a possible direction for a future research work.

% \bibliography{main}
\end{document}